\documentclass[10pt]{article}
\usepackage[utf8]{inputenc}
\usepackage{graphicx}
\usepackage{setspace}
\usepackage[\ifnum\pdfoutput=1breaklinks\fi]{hyperref}
\PassOptionsToPackage{margin=1in}{geometry}
\usepackage{epstopdf}
\usepackage{wordlike}
\usepackage{lipsum}
\usepackage{setspace}
\usepackage{amssymb,amsmath}
\usepackage{amsmath}
\usepackage{lineno}
\usepackage{url}
\usepackage{natbib}
\usepackage{graphicx}
\usepackage{fancyhdr}
 \usepackage{authblk}
\begin{document}


\title{Decoding the Encoding of Functional Brain Networks: an fMRI Classification Comparison of Non-negative Matrix Factorization (NMF), Independent Component Analysis (ICA), and Sparse Coding Algorithms}

\author[1]{Jianwen Xie}
\author[2]{Pamela K. Douglas}
\author[1]{Ying Nian Wu}
\author[2]{Arthur L. Brody}
\author[2]{Ariana E. Anderson \thanks{Department of Psychiatry and Biobehavioral Sciences, University of California, Los Angeles, 760 Westwood Plaza, Los Angeles, CA 90095 \\arianaanderson@mednet.ucla.edu} }
\affil[1]{Department of Statistics, University of California, Los Angeles}
\affil[2]{Department of Psychiatry and Biobehavioral Sciences, University of California, Los Angeles}
\renewcommand\Authands{ and }
\date{}
\newpage
\cleardoublepage
\clearpage

\maketitle
\newpage
\begin{abstract} 

Brain networks in fMRI are typically identified using spatial independent component analysis (ICA), yet mathematical constraints such as sparse coding and positivity both provide alternate biologically-plausible frameworks for generating brain networks.  Non-negative Matrix Factorization (NMF) would suppress negative BOLD signal by enforcing positivity. Spatial sparse coding algorithms ($L1$ Regularized Learning and K-SVD) would impose local specialization and a discouragement of multitasking, where the total observed activity in a single voxel originates from a restricted number of possible brain networks.

The assumptions of independence, positivity, and sparsity to encode task-related brain networks are compared; the resulting brain networks for different constraints are used as basis functions to encode the observed functional activity at a given time point. These encodings are decoded using machine learning to compare both the algorithms and their assumptions, using the time series weights to predict whether a subject is viewing a video, listening to an audio cue, or at rest, in 304 fMRI scans from 51 subjects. 

For classifying cognitive activity, the sparse coding algorithm of $L1$ Regularized Learning consistently outperformed 4 variations of ICA across different numbers of networks and noise levels (p$<$0.001).  The NMF algorithms, which suppressed negative BOLD signal, had the poorest accuracy. Within each algorithm, encodings using sparser spatial networks (containing more zero-valued voxels) had higher classification accuracy (p$<$0.001). The success of sparse coding algorithms may suggest that algorithms which enforce sparse coding, discourage multitasking, and promote local specialization may capture better the underlying source processes than those which allow inexhaustible local processes such as ICA. 

\end{abstract}

%
\newpage
\cleardoublepage
\clearpage
\section{Introduction}

Although functional MRI (fMRI) data contain numerous spatio-temporal observations, these data putatively reflect changes in a relatively small number of functional networks in the brain. These underlying processes can be modeled using any number of blind source separation (BSS) algorithms, with Independent Component Analysis (ICA) being the most commonly used.   Other mathematical constraints, such as positivity or sparsity, provide alternative interpretations for how the brain generates and encodes functional brain networks.  Comparing how BSS algorithms capture task-related activations will simultaneously compare these interpretations of functional encoding.  

The observed fMRI volume at a given time instance can be modeled as a linear combination of weighted spatial networks $X$ according to $y = dX$. This linear model encodes an fMRI scan, where the spatial maps correspond to a brain network, and the time series weights describe the contribution of that spatial map to the total functional activity observed at a given time. This linear model can be computed using a variety of BSS methods including: ICA, K-Means Singular Value Decomposition (K-SVD), $L1$ Regularized Learning, and Non-negative Matrix Factorization (NMF). Each of these methods applies different constraints to recover and numerically unmix sources of activity, leading to different representations of functional brain networks (spatial maps) and time series weights.   

ICA finds components that are maximally statistically independent in the spatial or temporal domain. Since the number of time points is typically less than the number of voxels, spatial ICA in particular has become the most studied approach for extracting and summarizing brain activity networks in fMRI (\cite{mckeown1997analysis}), where many of the networks computed by ICA correspond to previously identified large scale brain networks (\cite{smith2009correspondence}). In spatial ICA, maximizing spatial independence suggests that the ability of a voxel to contribute to a given brain network is not contingent on how it contributes to any other brain network, or to how many other brain networks it contributes; there is no bound to how strongly a voxel can contribute to a network since arbitrarily large positive and negative ``activations" are permitted across networks (\cite{calhoun2004independent}).  This permits local activations to be inexhaustible; no region is disqualified from participating in any brain network simply because it already participates strongly in another network. The validity of the spatial and temporal independence assumptions in fMRI have been the subject of lively debate for some time (\cite{friston1998modes};\cite{mckeown2003independent};\cite{daubechies2009independent};\cite{calhoun2013}). The most frequently-used ICA algorithms maximize independence by maximizing the non-Gaussianity (Fast ICA)  (\cite{hyvarinen1997fast}) or minimizing the mutual information (InfoMax) (\cite{bell1995information}). Recently ProDenICA has been shown to perform better on resting-state fMRI data (\cite{risk2013evaluation}).  

K-SVD (sparse dictionary learning) has been used to nominate networks both on the voxel (\cite{lee2010statistical};\cite{lee2011data};\cite{abolghasemi2013fast}) and region of interest (ROI) scale (\cite{eavani2012sparse}), and restricts many of the voxels in a network to have null (zero) values by limiting component membership of each voxel.  This biologically plausible constraint prohibits multitasking of a voxel, as no voxel can contribute to all processes simultaneously (\cite{spratling2014classification}).  More recently it was applied in simulated functional connectivity (FC) analyses to recover true, underlying FC patterns (\cite{leonardi2014disentangling}), where dynamic FC was found to be better described during periods of task by alternating sparser FC states.  Similarly, $L1$ Regularized Learning (LASSO) was applied in Parkinson's Disease to analyze fMRI functional connectivity during resting state (\cite{liu2015sticky}). Non-negative matrix factorization has been applied to fMRI data (\cite{wang2004detecting};\cite{potluru2008group};\cite{ferdowsi2010constrained};\cite{ferdowsi2011new}), where the alternating least squares NMF algorithm has been found superior to detect task-related activation compared to three other NMF algorithms (\cite{ding2012performance}). Imposing non-negativity in NMF suppresses all negative BOLD signal, while the parts-based representation which results from this non-negativity suggests that a subset of local-circuits may be a better representation of functional activity than geographically-distributed networks.  Previously, we used NMF to identify multimodal MRI, fMRI, and phenotypic profiles of Attention Deficit Hyperactivity Disorder (ADHD) (\cite{anderson2013non}).  Although the biological interpretations of these algorithms are not mutually exclusive to ICA, the success of a specific dictionary learning method may prioritize theories of encoding.  

In this paper, we evaluate which of the individual subject representations computed by ICA, K-SVD, NMF, and $L1$ Regularized Learning best encode task-related activations that are pertinent for classification, by embedding these components as features within a decoding framework. These representations can be compared by using the time series weights of the spatial maps for task classification, where each time point in an fMRI scan is encoded using the functional brain networks proposed by each algorithm.  Using the time series weights for the encodings, we predict which task a subject was performing during a scan by employing support vector machines (SVM) (\cite{burges1998tutorial}) and random forests (\cite{breiman2001random}).  We have recently leveraged component feature weights for classification of fMRI data in a number of studies (\cite{douglas2009naive};\cite{anderson2010classification};\cite{douglas2011performance};\cite{anderson2012real};\cite{douglas2013single}). We compare the predictive accuracies of the algorithms while varying the number of components and the presence of artifactual components (effects of motion, non-neuronal physiology, scanner artifacts and other noisy sources). Finally, we evaluate how physiological profiles of the proposed brain networks (tissue activation densities and sparsity) are associated with the classification accuracy, to compare whether the algorithms are substantially different after accounting for the physiological profiles of the spatial networks they extract.   Collectively, this paper evaluates the performance of different algorithms in encoding functional brain networks, possible correlates for explaining their performance, and the assumptions they support.

\section{Materials and Methods}

\subsection{Overview}

We will compare the representations of each algorithm by using the time series weights to predict, within a single scan, which activity a subject was doing.  We evaluate not only the general algorithms, but also their varied implementations, including four variations of ICA (Entropy Bound Minimization [EBM ICA], Fast ICA, InfoMax ICA, Joint Approximate Diagonalization of Eigen-matrices [JADE ICA]), two variations of NMF (Alternating Least Squares [NMF-ALS], Projected Gradient [NMF-PG]), and two sparse coding algorithms ($L1$ Regularized Learning, K-SVD).   Finally, we assess how the physiological profiles of the spatial maps may correlate with the ability to encode an fMRI scan, holding constant the effect of the algorithm.

\subsection{Data:  Design, Experiment, Preprocessing and Cleaning}
We describe briefly the experimental design and experiment here; it is discussed in detail in (\cite{culbertson2011effect}).  A total of 51 subjects were scanned in a study on craving and addiction. The subjects were divided into three groups, and scanned up to 3 times before and after treatment (with bupropion, placebo, or counseling) while watching a video and receiving audio cues meant to induce nicotine cravings, in a blocked design task.  This led to a total of 304 usable scans, after removing 2 scans for which scan-time was abbreviated.  There were a total of 18 nicotine related video cues, and 9 neutral video cues.  The audio cues were reportedly difficult to hear at times due to scanner noise. These volunteers were scanned using a gradient-echo, echo planar imaging sequence with a TR of 2.5 algorithms,  echo time, 45 milliseconds; flip angle, 80; image matrix, 128  64; field of view, 40  20 cm; and in-plane resolution, 3.125 mm.  

The fMRI data processing was carried out using FEAT (FMRI Expert Analysis Tool) version 6.00, part of FSL (FMRIB's Software Library, \url{www.fmrib.ox.ac.uk/fsl)}. The following preprocessing was applied; motion correction using MCFLIRT (\cite{jenkinson2002improved}); non-brain removal using BET (\cite{smith2002fast}); spatial smoothing using a Gaussian kernel of FWHM 5mm; grand-mean intensity normalisation of the entire 4D dataset by a single multiplicative factor; highpass temporal filtering (Gaussian-weighted least-squares straight line fitting, with sigma=50.0s).

We evaluated the classification accuracy of these algorithms on scans containing two levels of noise: the first dataset was traditionally-preprocessed using only the steps listed above, while the second dataset consisted of the same set of scans which had been additionally cleaned using the FIX artifact removal tool (\cite{griffanti2014ica};\cite{salimi2014automatic}), where approximately 50\% of networks (defined by ICA) were flagged as possible noise (residual effects of motion, non-neuronal physiology, scanner artifacts and other noisy sources) and removed.  Following the removal of the possible artifacts, each scan was reconstructed, and the dictionary learning algorithms were rerun within each ``cleaned" scan.  

\subsection{BSS Algorithms}

K-SVD, NMF, ICA, and $L1$ Regularized Learning all perform a matrix factorization into spatial maps (networks) and time series weights.  They differ primarily however in what constraints they impose when learning, and whether the primary emphasis is on learning the spatial or temporal features.  For example, NMF places equal emphasis on learning the time series weights and the networks, while spatial ICA imposes statistical independence over space (but no constraints over time).  In addition, the sparse coding algorithms considered here emphasize learning the time series weights instead of the spatial maps, and imposes spatial sparsity within voxel across networks by restricting the total contribution of a given voxel over all networks. Because all the algorithms assessed here produce both spatial maps and time courses, this distinction is not restrictive in comparing the ability of spatial maps to summarize functional activation patterns at a given time point. We uniformly describe the algorithms in the context of $Y=DX$, where $Y$ is the original data matrix (a single fMRI scan) of size $n \times m$ containing $m$ voxels and $n$ time points, $D$ is the mixing matrix containing the time series weights for $k$ components of dimension $n \times k$ where $k \leq n$, and $X$ is the matrix of spatial maps (networks) of dimension $k \times m$.  We extract both $k \in {20,50}$ networks within each scan for ICA, NMF, K-SVD and $L1$ Regularized Learning.


\subsubsection{Spatial Statistical Independence: Independent Component Analysis}

Statistical independence is an important concept that constitutes the foundation of ICA, which is one of the most widely used blind source separation techniques for exposing hidden factors underlying sets of signals (\cite{Aapo2000ICA};\cite{Aapo2001ICA}).  ICA can be written as a decomposition of a data matrix $Y_{n \times m}$ into a product of (maximally) statistically independent spatial maps (networks) $X_{k \times m}$ with a mixing matrix (time series weights) $D_{n \times k}$, given by  $Y_{n \times m}=D_{n \times k} X_{k \times m}$.  Formally, spatial independence implies that for a given voxel $m$, $p(x_{1m},x_{2m},\ldots, x_{km} )= \prod_{i=1}^k  p(x_{1m}) p(x_{2m})\ldots p(x_{km})$  Here, we examine ICA with a constraint of (maximal) spatial independence, using the Fast ICA, InfoMax ICA, EBM ICA, and JADE ICA algorithms. 

Different measurements of independence govern different forms of the ICA algorithms, resulting in slightly different unmixing matrices. Minimization of mutual information and maximization of non-Gaussianity are two broadest measurements of statistical independence for ICA. Fast ICA (\cite{Aapo1999ICA}) is a fixed point ICA algorithm that maximizes non-Gaussianity as a measure of statistical independence, motivated by the central limit theorem. Fast ICA measures non-Gaussianity by negentropy, which itself is the difference in entropy between a given distribution and the Gaussian distribution with the same mean and variance. InfoMax ICA (\cite{bell1995information}) belongs to the minimization of mutual information family of ICA algorithms; these find independent signals by maximizing entropy. Instead of directly estimating the entropy, the EBM ICA algorithm (\cite{li2010independent}) approximates the entropy by bounding the entropy of estimates using numerical computation.  Due to the flexibility of the entropy bound estimator, EBM ICA can be applied to data that come from different types of distributions, e.g., sub- or super-Gaussian, unimodal or multimodal, symmetric or skewed probability density functions, placing an even stronger emphasis on independence. JADE (\cite{cardoso1999high};\cite{cardoso1993blind}) is an ICA algorithm exploiting the Jacobi technique to perform joint approximate diagonalization on fourth-order cumulant matrices to separate the source signals from mixed signals. Typical ICA algorithms use centering, whitening, and dimensionality reduction as preprocessing steps. Whitening and dimension reduction can be achieved with Principal Component Analysis (PCA) or Singular Value Decomposition (SVD). They are simple and efficient operations that significantly reduce the computational complexity of ICA, so are applied in the implementations of ICA here.

\subsubsection{Positivity: Non-negative Matrix Factorization}
Psychological and physiological evidence show that component parts play an important role in neural encoding of the visual appearance of an object in the brain (\cite{palmer1977hierarchical};\cite{logothetis1996visual};\cite{wachsmuth1994recognition}).   When applied to fMRI data, this parts-based representation is conceptually similar to encouraging neighboring voxels to co-activate, which would encourage spatial smoothness in the resulting component maps.  The non-negativity in NMF constraints would eliminate negative BOLD signal changes which have been controversially associated with sources such as cerebral blood volume changes and inhibition (\cite{bianciardi2011negative};\cite{harel2002origin};\cite{moraschi2012origin};\cite{smith2004negative}).

Non-negativity is a useful constraint for matrix factorization which leads to parts-based representation, because it allows only additive (positive) combinations of the learned bases (\cite{Lee1999NMF}). Given a $n \times m$ non-negative data matrix $Y$ and a pre-specified $k< \min(n,m)$, NMF finds two non-negative matrices $D_{n \times k}$ and $X_{k \times m}$ such that $Y \approx DX$. The conventional approach to find $D$ and $X$ is by minimizing the squared error between $Y$ and $DX$:

\begin{eqnarray}\label{ab7}
\min_{D,X}\{\left\Vert Y-DX\right\Vert _{2}^{2}\}
\mathrm{subject\: to}D_{ia}\geq0,X_{bj}\geq0,\forall i,a,b,j
\end{eqnarray}
which is a standard bound-constrained optimization problem. There are several algorithms in which the $D$ and $X$ may be found. The most commonly known approach is the multiplicative update method (\cite{Lee2001NMF}). NMF-ALS has previously shown stronger performance than multiplicative update-NMF in fMRI (\cite{ding2012performance}).
 
\textbf{NMF-ALS}

 A more flexible and general framework for obtaining NMF is to use alternating least squares (ALS), which was first introduced by Paatero and Tapper (\cite{Paatero1994NMF}) in the middle of the 1990s under the name positive matrix factorization (\cite{anttila1995source};\cite{Paatero1994NMF}). The ALS algorithm does not have the restriction of locking 0 elements in matrices $D$ and $X$. The framework of ALS is summarized as follows:

(1) Initialize $D_{ia}^0>0, X_{bj}^0>0, \forall i,a,b,j$.

(2) For $J=0,1,2, \cdots$
\begin{eqnarray} \label{ab7} 
 D^{(J+1)}= \arg \min_{D\geq 0} \bigl \Vert Y-DX^{(J)} \bigr \Vert_{2}^{2} \label{ANLS:a}\\
 X^{(J+1)}= \arg \min_{X\geq 0} \bigl \Vert Y-D^{(J+1)}X \bigr \Vert_{2}^{2}  \label{ANLS:b}
\end{eqnarray}

The iterations can be performed with an initialization of $D$ and $X$, and then alternating between solving (\ref{ANLS:a}) and (\ref{ANLS:b}) until a stopping criterion is satisfied. ALS is also known as "block coordinate descent" approach in bound-constrained optimization (\cite{dimitri1999NonlinearProgramming}). We refer to (\ref{ANLS:a}) or (\ref{ANLS:b}) as a sub-problem in this.  At each step of iterations, finding an optimal solution of the nonnegative least squares sub-problem is important because otherwise, the convergence of overall algorithm may not be guaranteed (\cite{kim2007NMF}). 

Some successful NMF algorithms are based on ALS; their differences arise from using different ways to solve the ALS sub-problems. As an elementary strategy, the alternating least squares algorithm (\cite{Paatero1994NMF};\cite{Berry2007NMF}) solves the sub-problems by an unconstrained least squares solution (without the nonnegativity constraint), \textit{i.e.}, $D \leftarrow ((XX^T)^{-1}XY^T)^T$ and $X \leftarrow (D^TD)^{-1}D^TY$, and every negative element resulting from least squares computation is set to zero to ensure nonnegativity after each update step. The implementation for ALS described above is very fast, and requires less work than other NMF algorithms; however, setting negative elements to 0 in order to enforce nonnegativity is quite \textit{ad hoc}.

\textbf{NMF-PG}

Alternating nonnegative least squares using projected gradients (NMF-PG) has been used for NMF (\cite{Lin2007NMF}). The sub-problems in ALS above are solved here using projected gradient methods. To calculate $X$, the algorithm updates it by the rule $X \leftarrow P[X- \alpha \nabla f(X)]$, where $P[ \cdot ]$ is a bounding function or projection operator that maps a point back to the bounded feasible region, $\nabla f(X)$ is the gradient function computed as $D^T(DX-Y)$, and $\alpha$ is the step size. Selecting the step size $\alpha$ for each sub-iteration in NMF-PG is a main computational task. The same approach is utilized to calculate $D$. 

\subsubsection{Sparse coding: K-SVD and L1 Regularized Learning}

All dimension reduction methods necessarily provide compression, and some variations of independence also encourage sparsity (\textit{e.g.} the InfoMax variant of ICA).  Similarly, the non-negativity constraint in NMF shrinks the intensity of many spatial maps' voxels to zero which also encourages sparsity.  However, the sparsity obtained in these methods is a secondary benefit to the primary intention (independence and non-negativity), and not the primary objective of the algorithms.  We thus describe the ``sparse coding" algorithms not on whether they may encourage sparsity, but rather on whether they enforce it.  Sparse coding in fMRI restricts a single voxel to have a limited number of (approximately) non-null values across networks. This sparsity on the voxel scale across spatial maps necessarily provides sparsity within each spatial map as well.

\textbf{K-SVD}\\
K-SVD is a generalization of the k-means algorithm; in the k-means algorithm, an observation can be represented by its centroid (the central point of the cluster to which that element belongs).  In K-SVD, an observation is instead modeled as a weighted combination of multiple (but not all) dictionary elements- effectively imposing the $L0$-norm on how many networks a specific voxel can participate in.   K-SVD is a sparse data-representation algorithm to learn an over-complete dictionary, such that any single observation is constructed from a subset of the total number of dictionary elements.   

Sparsity is enforced over space by limiting the number of elements which can be used to construct that observation, and the dictionary elements $D$ learned are the corresponding time series weights; when applied to fMRI, this constraint more generally suggests a local specialization; a single voxel can only contribute to a subset of all ongoing processes. The weights of the dictionary (time series weights) are the spatial maps themselves.  This bypasses the need for the PCA which typically precedes ICA.  The sparse coding constraint over space suggests that the networks are best represented by a subset of all voxels; this is in direct contrast to algorithms such as ICA, where every voxel is allowed to contribute, in varying degrees, to the representative time series weights.

K-SVD operates by iteratively alternating between (1) sparse coding of the data based on the current dictionary (estimating the spatial maps, when applied to fMRI), and (2) dictionary updating (revising the time series weights) to better fit the observed data (\cite{Elad2006Ksvd};\cite{rubinstein2010dictionaries}). A full description of the K-SVD algorithm is given as follows:

Task: Find the best dictionary to represent the data samples $\{y_{i}\}$ as sparse compositions, by solving
\begin{equation}\label{ab7}
\min_{D,X}\{\left\Vert Y-DX\right\Vert _{2}^{2}\}\mathrm{subject\: to}\forall i,\left\Vert x_{i}\right\Vert_{0} \leq T_{o}
\end{equation}

Initialization: Set the dictionary matrix $D^{(0)}\in R^{n \times K}$ with $l^{2}$ normalized columns. Set $J=0$, the counting index.  Let $n$ = the number time points, $m$= the number of voxels, and let $K$ = the number of dictionary elements being estimated.  

Main Iteration: Increment $J$ by 1, and apply:

\underline{Sparse Coding Stage:} Use any pursuit algorithm to compute the representation vectors $x_i$ for each example $y_i$, by approximating the solution of
\begin{equation}\label{ab7}
i=1,2,...,m,\min_{x_{i}}\{\bigl\Vert y_{i}-D^{(J-1)}x_{i}\bigr\Vert_{2}^{2}\}\mathrm{ subject\: to}\left\Vert x_{i}\right\Vert _{0}\leq T_{0}
\end{equation}

\underline{Dictionary Update Stage}: for each column $k=1,2,\cdots,K$ in $D^{(J-1)}$, update it as follows:

 (1) Define the group of observations that use this atom, $\omega_{k}=\{i\mid1\leq i\leq m,x_{T}^{k}(i)\neq0\}$.
 
 (2) Compute the overall representation error matrix, $E_k$, by $E_{k}=Y-\sum_{j\neq k}d_{j}x_{T}^{j}$.
 
 (3) Restrict $E_k$ by choosing only the columns corresponding to $\omega_{k}$, and obtain $E_k^R$.
 
 (4) Apply SVD decomposition $E_k^R=U \Delta V^T$. Choose the updated dictionary column $d_k$ to be the first column of $U$. Update the coefficient vector $x_R^k$ to be the first column of $V$ multiplied by $\Delta(1,1)$.

Stopping rule: If the change in $\bigl\Vert Y-D^{(J)}X^{(J)}\bigr\Vert_{2}^{2}$ is small enough, stop. Otherwise, iterate further.

Output: The desired results are dictionary $D^{J}$ and encoding $X^{J}$.

Due to the $L0$-norm constraint, seeking an appropriate dictionary for the data is a non-convex problem, so K-SVD does not guarantee to find the global optimum( \cite{rubinstein2010dictionaries}).

\textbf{L1 Regularized Learning}\\
We can relax the $L0$-norm constraint over the coefficients $x_i$ by instead using a $L1$-norm regularization (\cite{olshausen1996emergence}), which enforce $x_i$ $(i = 1,...m)$ to have a small number of nonzero elements. Then, the optimization problem can be written as:

\begin{equation}\label{ab7}
\min_{D,X}\{\left\Vert Y-DX\right\Vert _{2}^{2}\}+\beta\sum_{i}\left\Vert x_{i}\right\Vert _{1}
\mathrm{subject\: to}\left\Vert d_{j}\right\Vert _{2}\leq1,\forall j=1,2,\cdots,k
\end{equation}
where a unit $L2$-norm constraint on $d_j$ typically is applied to avoid trivial solutions.

Due to the use of $L1$ penalty as the sparsity function, the optimization problem is convex in $D$ (while fixing $X$) and convex in $X$ (while fixing $D$), but not convex in both simultaneously. \cite{Lee07sparseCoding} optimizes the above objective iteratively by alternatingly optimizing with respect to $D$ (dictionary) and $X$ (coefficients) while fixing the other. For learning the coefficients $X$, the optimization problem can be solved by fixing $D$ and optimizing over each coefficient $x_i$ individually:
\begin{equation}\label{ab7}
\min_{x_i}\{\left\Vert y_{i}-Dx_{i}\right\Vert _{2}^{2}\}+\beta\left\Vert x_{i}\right\Vert _{1}
\end{equation}
which is equivalent to $L1$-regularized least squares problem, also known as the Lasso in statistical literature. For learning the dictionary $D$, the problem reduces to a least square problem with quadratic constraints:

\begin{equation}\label{ab7}
\min_{D} \left\Vert Y-DX\right\Vert _{2}^{2}
\mathrm{subject\: to}\left\Vert d_{j}\right\Vert _{2}\leq1,\forall j=1,2,\cdots,k
\end{equation}
\subsection{Implementation Details: BSS Algorithms}

Here, we have tried to explore common variations for each learning algorithm as thoroughly as is computationally feasible. Full implementation code is provided in the appendix, with a summary of implementation provided here. This section describes the implementation and the crucial parameters used in each learning algorithm.  Given an input matrix $Y$, all the algorithms were initialized by randomly generating matrix $D$ and $X$, and run a sufficiently large number of iterations to obtain the converged results. For most of the algorithms, the number of iterations we used is 400, upon which we verified convergence using appropriate fit indices.  

NMF: We used Matlab's embedded function {\em nnmf} for NMF-ALS and (\cite{Lin2007NMF}) for NMF-PG in our experiment. Maximum number of 400 iterations is allowed. 
  
InfoMax ICA: We used the EEGLAB toolbox (\cite{EEGLAB}) for InfoMax ICA, which implements logistic InfoMax ICA algorithm of Bell and Sejnowski (\cite{bell1995information}) with the natural gradient feature (\cite{amari1996new}), and with PCA dimension reduction. Annealing based on weight changes is used to automate the separation process. The algorithm stops training if weights change below $10^{-6}$ or after 500 iterations.

Fast ICA: We used the Fast ICA package (\cite{fastICAsoftware}), which implements the fast fixed-point algorithm for ICA and projection pursuit. PCA dimension reduction and hyperbolic tangent for nonlinearity are used.

JADE ICA: We used the Matlab implementation of JADE (\cite{cardoso1993blind}) for the ICA of real-valued data.  PCA is used for dimension reduction before the JADE algorithm is performed.

EBM ICA: We used the Matlab implementation of EBM ICA (\cite{li2010independent}) for real-valued data. Four nonlinearities (measuring functions) $x^4$,  $|x|/(1+|x|)$, $x|x|/(10+|x|)$, and $x/(1+x^2)$ are used for entropy bound calculation. This implementation adopts a two-stage procedure, where the orthogonal version of EBM ICA, with measuring function $x^4$ and maximum number of iterations of 100, is firstly used to provide an initial guess, and then the general nonorthogonal EBM ICA with all measuring functions uses the linear search algorithm to estimate the demixing matrix. The technique for detection and removal of saddle convergence proposed in (\cite{koldovsky2006efficient}) is used in orthogonal EBM ICA if the algorithm converges to a saddle point. Similar to other ICA methods, PCA for dimension reduction was used before the algorithm is performed. 

K-SVD: While the total number of components was allowed to vary (either 20 or 50), each voxel was allowed to participate in only K-8 networks. This corresponds to 40\% of all components for the 20-network extractions, and 16\% of components for the 50-network extraction. The K-SVD-Box package (\cite{Rubinstein2008Ksvd}) was used to perform K-SVD, which reduces both the computing complexity and the memory requirements by using a modified dictionary update step that replaces the explicit SVD computation with a much quicker approximation. It employs the Batch-OMP (orthogonal matching pursuit) to accelerate the sparse-coding step. Implementation details can be found in (\cite{Rubinstein2008Ksvd}).  

$L1$ Regularized Learning: We used the implementation of efficient sparse coding proposed by \cite{Lee07sparseCoding}. It solves the $L1$-regularized least squares problem iteratively by using the feature-sign search algorithm and $L2$-constrained least squares problem by its Lagrange dual. The parameter for sparsity regularization is 0.15 and for smoothing regularization is $10^{-5}$.

\subsection{Implementation Details: Machine Learning Algorithms and Parameters}

Using the time series weights for each algorithm extracted within each scan, we predicted whether a subject was viewing a video, resting, or listening to an audio stimulus at each time point.  This was done separately for both 20 and 50 network extractions, and for both the traditionally-preprocessed and artifact-suppressed data, to assess the impact of both network number and the effect of residual noise. The most stringent data cleaning involved cleaning the scans twice, using traditional pre-processing and FIX where artifactual components were identified and discarded from the scan (residual effects of motion, non-neuronal physiology, scanner artifacts and other noisy sources).  We predicted which activity a subject was performing using both an SVM classifier (using a 10-fold cross-validation) as well as a random forests classifier (which provides the out-of-bag testing error).  Using the time series weights for prediction is similar to projecting the entire fMRI scan onto the spatial networks defined by the algorithms. The average classification accuracy over 304 scans measures the predictive performance of each algorithm, for the specified number of components and artifact suppression level.

SVM with a radial basis kernel was implemented within R (\cite{e1071}), using default parameter settings (cost parameter: 1,  gamma:  0.05).    For multiclass-classification with 3 levels as was implemented here (video, audio, rest), libsvm (called from R) uses the ``one-against-one'' approach, in which 3 binary classifiers are trained; the appropriate class is found by a voting scheme. Random Forests was implemented within R with 500 trees using default parameter settings (\cite{liaw2002classification}).  For the 20 networks, 4 variables were tried at each split.  For the 50 networks, 7 variables were tried.

\subsection{Measuring Noise and Measuring Sparsity within Extracted Networks}

After performing two iterations of data cleaning (traditional preprocessing and FIX artifact removal), we subsequently measured whether residual noise or sparsity may impact the classification accuracy.   We hypothesized that ``activation'', or high-intensity voxel values, within CSF regions may be an indicator of the overall level of noise within a spatial map.  For the spatial maps created by running each BSS algorithm on the artifact-cleaned scans, we measured the average intensity of voxels within CSF, grey, and white matter regions. The T1 MNI tissue partial volume effect (PVEs) were aligned into the subject's functional space \text{via} the subject's T2 structural scan in a two-step process.  First, the segmented MNI images were aligned into the subject's structural space using the whole-brain MNI-152 to the subject's T2 mapping learned using FLIRT.  Then, we registered these PVE images into the subject's functional MRI space using the subject's T2 to fMRI mapping. 

Using these tissue masks we computed the average intensity of the extracted spatial maps within regions probabilistically defined as grey matter, white matter, and CSF. This was computed using the cross-correlation of each tissue-type partial volume effect (PVE) with each functional map; for a given algorithm, the 20 networks extracted for a scan would yield 60 correlation measures with the grey, white, and CSF maps.  The average and the variation of the tissue types in the 20 networks were used to summarize the overall distribution of ``active" voxels in the spatial maps. These tissue-region correlates were computed for the 2,432 basis sets extracted for all algorithms.  

To measure sparsity for each BSS algorithm within each scan, we computed the $L0$-norm of each spatial map, and used the average across all networks within a scan to measure the spatial sparsity of the extracted networks.  Specifically, for each spatial map we measured sparsity using the $L0$-norm, where
$sparsity(X)=\frac{-1}{k}*\sum_{j=1}^k \Vert X_{j } \Vert_{0} $ where $k$ is the total number of components.  The negative sign ensures that more zero-valued voxels will lead to a lower sparsity measure.  This $L0$ based measure was chosen because it is not sensitive to the scaling of the images, which are necessarily different across algorithms.

\subsection{Comparison of BSS Algorithms by Classification Accuracy}
The BSS algorithms' SVM performance were first compared for the 20 network, artifact-cleaned scans, predicting accuracy using Algorithm as a main effect in a general linear mixed-effects regression model (baseline model); Scan-ID and Subject-ID were included as random effects to adjust for multiple comparisons. Across the 304 scans and 8 algorithms, this resulted in the classification accuracy from 2,432 SVM models being compared.  We then assessed whether sparser spatial maps led to better classification accuracy after adjusting for the effect of the algorithm, predicting classification accuracy using both the BSS algorithm type and network sparsity as fixed effects and Scan-ID and Subject-ID as random effects for multiple comparison adjustments.  

Finally, we included tissue-type profiles to measure the association of the physiological profiles of the spatial maps with their ability to decode functional activity.  This assesses, among other things, whether network extractions containing larger amounts of white matter have poorer classification accuracy than network extractions containing more grey matter.  In a general linear mixed effects regression model (full model), we predicted the classification accuracy within each scan using the algorithm type, a session effect, the sparsity of the networks, the average and standard deviation of the intensity within regions of grey matter, white matter, CSF (averaged across networks).  Scan ID and Subject ID were both included as random effects to adjust for multiple comparisons.  This was done for the 20 network extraction, on data which had been cleaned of artifacts (including white matter and CSF artifacts).  

We compared the baseline model containing just the BSS algorithm main effect to the full model containing the BSS algorithm effect and the physiological profiles of the spatial maps using a chi-square hierarchical regression, to evaluate whether the physiological profiles of the spatial maps were related to classification accuracy above and beyond the effect of the BSS algorithm alone.

\section{Results}
\begin{figure}[h]
\begin{center}
\includegraphics[width=1.0\textwidth]{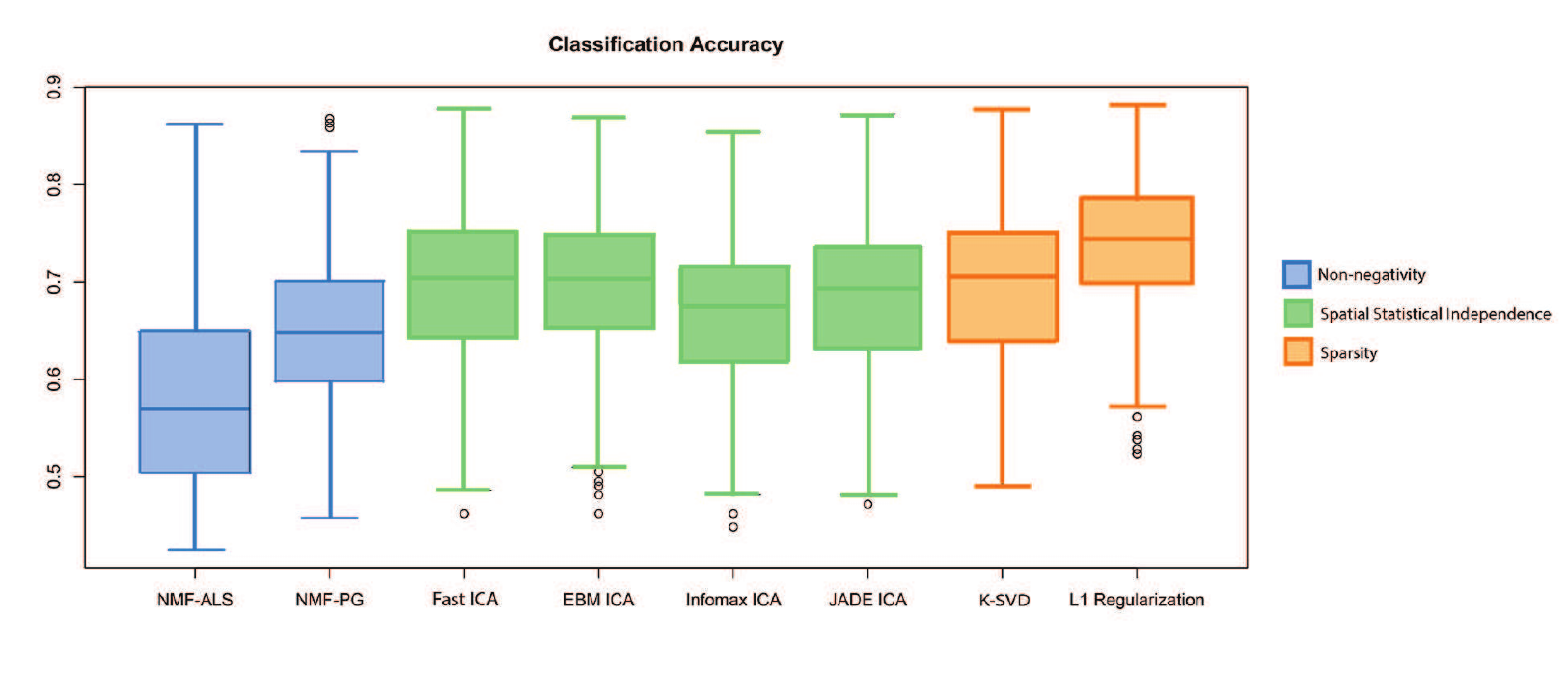}
\caption{Although ICA is the most commonly used method to extract and define fMRI spatial networks, encoding scans using sparse coding algorithms like K-SVD and $L1$ Regularized Learning led to higher classification accuracy in determining whether a subject was resting, watching a video, or hearing an audio cue.  The best performing independence algorithm (Fast ICA) was still inferior to the worst performing sparse coding algorithm (K-SVD) for classifying cognitive activity (p$<$0.005) using an SVM classifier on 20 networks on traditionally-preprocessed data.  Chance accuracy is 50\%.   \label{fig:results}}
\end{center}
\end{figure}

All algorithms performed better than chance on average, with general trends present across the constraint families.  The best performing independence algorithm (Fast ICA) was still inferior to the worst performing sparse coding algorithm (K-SVD) for classifying cognitive activity (p$<$0.005) in Table \ref{tab:tableresults20svm} and Figure \ref{fig:results}, using 20 dictionary elements on the traditionally-preprocessed fMRI data and an SVM classifier.  The strong predictive performance of the sparse coding algorithms persisted when using 50 networks instead of 20 (Table \ref{tab:tableresults50svm}) and when using data from which additional artifacts had been removed, shown in Table \ref{tab:tableresults20svmclean}.  

\begin{figure}[h]
\begin{center}
\includegraphics[width=1\textwidth]{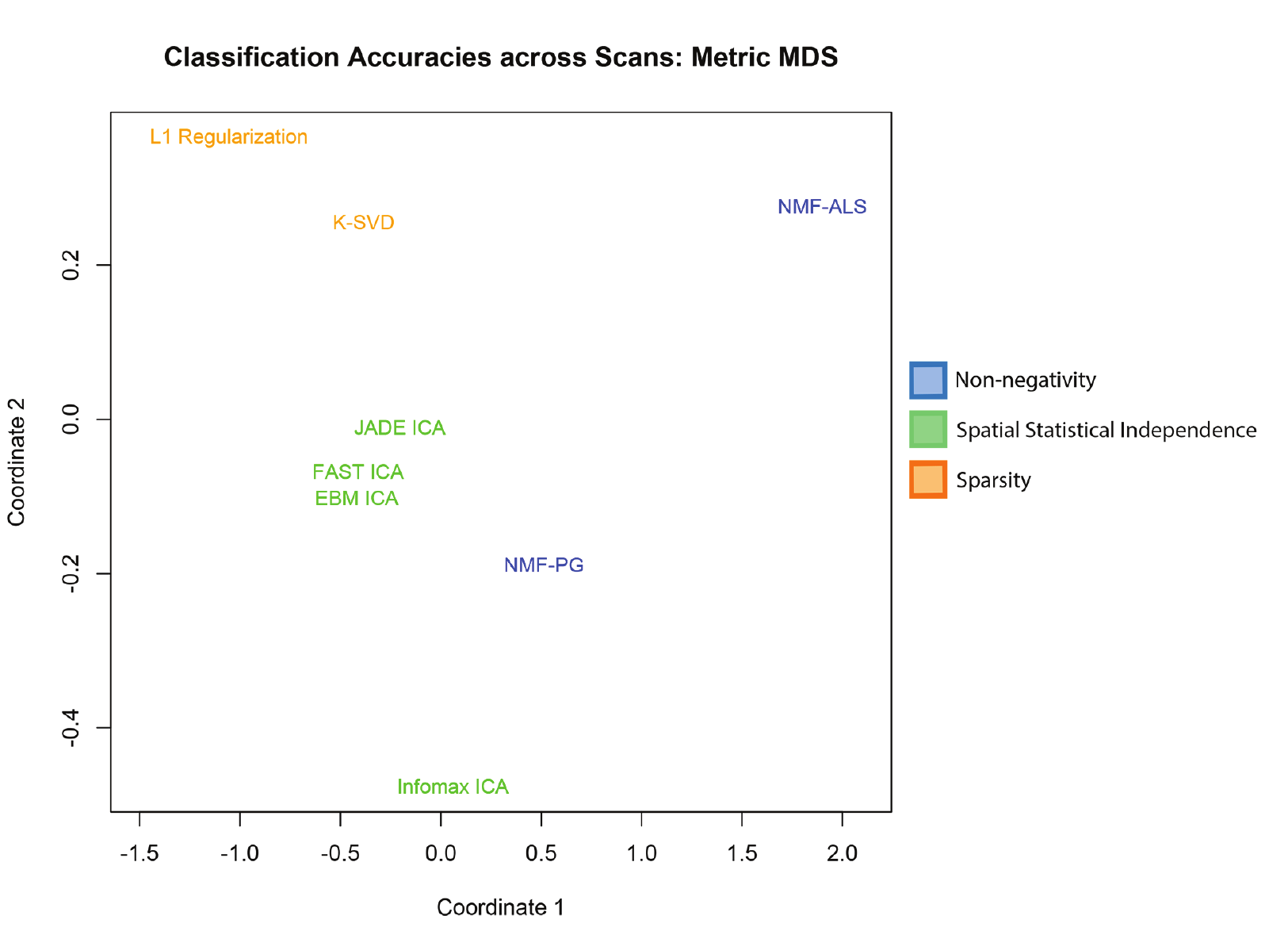}
\caption{Within a scan, algorithms sharing particular constraints tended to classify with similar accuracy.  This multi-dimensional scaling visualizes the similarities of how the BSS algorithms classified within each scan using the 20-basis networks extracted on the traditionally preprocessed data. \label{fig:mdsaccuracies}}
\end{center}
\end{figure}

We compared how similarly these algorithms classified within each scan in Figure \ref{fig:mdsaccuracies}, using a multi-dimensional scaling of the accuracies within scan (for the 20-network, traditionally preprocessed scans).  Methods predicted similarly to other methods within their class: K-SVD, a sparse coding method, was most correlated with $L1$ Regularized Learning (Lasso), another sparse coding method.  This suggests that algorithms within certain families tend to have similar performance.

Within each scan, the algorithms in order from best to worst classification accuracy were: $L1$ Regularization, K-SVD, Fast ICA,  EBM ICA, JADE ICA, InfoMax ICA, NMF-PG, and NMF-ALS, as shown in Table \ref{tab:nophysioclass} using 20 spatial maps on artifact-cleaned data.  Spatial sparsity was highly significant; scans which extracted sparser spatial maps had higher classification accuracy ($p<0.001$) after accounting for the effect of the algorithms as shown in Table  \ref{tab:sparsityalgclass}.  The CSF functional map density was negatively associated with classification accuracy ($p<0.001$), while high variability in white matter functional density was associated with better classification accuracy ($p<0.001$). Including the physiological profiles of the spatial maps significantly helped predict the within-scan classification accuracy, above and beyond the effect of the algorithm alone ($p<0.001$, chi-square test of nested models).  The physiological profiles of the spatial maps were likely correlated with sparsity of the spatial maps, as the sparsity measurement lost significance once accounting for the physiological profiles.

\begin{table}[ht]
\centering
\begin{tabular}{rrr}
  \hline
 & SVM & Random Forests  \\ 
  \hline
  NMF-ALS & 0.63 (0.08) & 0.63 (0.09) \\ 
  NMF-PG & 0.63 (0.08) & 0.63 (0.09)\\ 
  InfoMax ICA & 0.64 (0.08) & 0.67 (0.09)\\ 
  JADE ICA & 0.66 (0.08) & 0.69 (0.09) \\ 
  EBM ICA & 0.66 (0.10) & 0.71 (0.10) \\
  Fast ICA & 0.67 (0.08) & 0.72 (0.08)\\ 
  K-SVD & 0.70 (0.08) & 0.73 (0.08)  \\ 
  $L1$ Regularized Learning & 0.74 (0.07) & 0.71 (0.07) \\ 
   \hline
\end{tabular}
\caption{Classification accuracy averaged across 304 traditionally preprocessed data scans in predicting whether a subject was viewing a video, listening to an audio stimuli, or resting, using 20 networks.  Chance accuracy is 50\%.  \label{tab:tableresults20svm}}
\end{table}

\begin{table}[ht]
\centering
\begin{tabular}{rrr}
  \hline
 & SVM & Random Forests  \\ 
  \hline
InfoMax & 0.59 (0.07) & 0.59 (0.07) \\ 
  NMF-ALS & 0.60 (0.08) & 0.62 (0.10)\\ 
  NMF-PG & 0.66  (0.09) & 0.64 (0.08) \\ 
  JADE ICA & 0.70 (0.08) & 0.72 (0.08) \\ 
  EBM ICA&0.72 (0.09) & 0.74 (0.09) \\
  Fast ICA & 0.73 (0.07) & 0.75 (0.07) \\ 
  K-SVD & 0.75 (0.07) & 0.77 (0.07) \\ 
  $L1$ Regularized Learning & 0.75 (0.07) & 0.69 (0.07)\\ 
   \hline
\end{tabular}
\caption{Classification accuracy averaged across 304 traditionally preprocessed scans in predicting whether a subject was viewing a video, listening to an audio stimuli, or resting, using 50 networks. Chance accuracy is 50\%. \label{tab:tableresults50svm}} 
\end{table}

\begin{table}[ht]
\centering
\begin{tabular}{rrr}
  \hline
 & SVM & Random Forests  \\ 
  \hline
NMF-ALS & 0.58 (0.09) & 0.60 (0.11)\\ 
  NMF-PG & 0.58 (0.09)  &  0.59 (0.11)\\ 
  InfoMax ICA & 0.67 (0.07) & 0.67 (0.07) \\ 
  JADE ICA & 0.68 (0.08) & 0.71(0.08)\\ 
  EBM ICA & 0.69 (0.09) & 0.72 (0.09)\\ 
  Fast ICA & 0.70 (0.08) & 0.72 (0.08) \\ 
  K-SVD & 0.70 (0.08) & 0.71 (0.08)\\ 
     $L1$ Regularized Learning & 0.74 (0.07) & 0.71 (0.07) \\ 
   \hline
\end{tabular}
\caption{Classification accuracy averaged across 304 artifact-cleaned scans in predicting whether a subject was viewing a video, listening to an audio stimuli, or resting, using 20 networks. Chance accuracy is 50\%. \label{tab:tableresults20svmclean}} 
\end{table}
\begin{table}[ht]
\centering
\begin{tabular}{rrrl}
  \hline
 & Estimate & Std. Error & t value \\ 
  \hline
(Intercept) & 0.70 & 0.01 & 110.55 ***\\ 
  NMF-ALS& -0.11 & 0.001 & -37.84 ***\\ 
  NMF-PG & -0.05 & 0.001 & -15.59 ***\\ 
  InfoMax ICA & -0.03 & 0.001 & -10.11 ***\\ 
  JADE ICA& -0.01 & 0.001 & -4.02 ***\\ 
    EBM ICA& -0.001 & 0.001 & -0.91 \\ 
 K-SVD & 0.001 & 0.001 & 0.36 \\ 
  $L1$ Regularized Learning & 0.04 & 0.001 & 14.70 ***\\ 
   \hline
\end{tabular}\caption{Classification accuracy of BSS algorithm compared to Fast ICA, using 20 networks extracted from artifact-cleaned scans, in order of performance from worst to best.   Time series weights from InfoMax ICA, JADE ICA, NMF-PG, NMF-ALS predicted activity significantly worse than Fast ICA, while L1-Regularization did significantly better ($p<0.001$).  Baseline is set to Fast ICA classification accuracy.  Scan-ID and Subject-ID were included as random effects within a general linear mixed-effects regression model to adjust for multiple comparisons.  *** = p $<$ .001 \label{tab:nophysioclass}}
\end{table}
\begin{table}[ht]
\centering
\begin{tabular}{rrrl}
  \hline
 & Estimate & Std. Error & t value \\ 
  \hline
(Intercept) & 0.83 & 0.01 & 57.91 ***\\ 
Sparsity & 0.001 & 0.001 & 10.82 ***\\ 
NMF-ALS& -0.14 & 0.001 & -35.21 ***\\ 
 K-SVD & -0.08 & 0.01 & -9.99 *** \\ 
 NMF-PG & -0.05 & 0.001 & -17.48 ***\\ 
  InfoMax ICA & -0.03 & 0.001 & -10.39 ***\\ 
  JADE ICA& -0.01 & 0.001 & -4.13 ***\\ 
  EBM ICA& -0.001 & 0.001 & -0.93 \\ 
  $L1$ Regularized Learning & 0.03 & 0.001 & 10.70 **\\ 
   \hline
   \end{tabular}
\caption{Greater sparsity for an extracted spatial map was associated with a higher classification accuracy in predicting a subject's task during scan time when using those spatial maps for encoding ($p<0.001$), holding constant the effect of the algorithm. Using 20 networks extracted from artifact-cleaned scans, sparsity was measured using the negative averaged number of zero-valued voxels of all spatial maps, which is insensitive to the scaling of the individual algorithms.  Baseline is set to Fast ICA classification accuracy. Scan-ID and Subject-ID were included as random effects within a general linear mixed-effects regression model to adjust for multiple comparisons.  *** = p $<$ .001  \label{tab:sparsityalgclass}}
\end{table}
\begin{table}[ht]
\centering
\begin{tabular}{rrrl}
  \hline
 & Estimate & Std. Error & t value \\ 
  \hline
(Intercept) & 0.71 & 0.02 & 30.53 ***\\ 
    Mean(CSF) & -0.30 & 0.09 & -3.34 ***\\ 
  Mean(Grey Matter) & -0.05 & 0.08 & -0.65 \\ 
  Mean(White Matter) & 0.01 & 0.06 & 0.08 \\ 
  SD(CSF) & 0.001 & 0.11 & 0.01 \\ 
  SD(Grey Matter) & -0.32 & 0.09 & -3.62 ***\\ 
  SD(White Matter) & 0.44 & 0.07 & 6.07 ***\\ 
   Sparsity & 0.001 & 0.001 & 0.68 \\ 
  Session & -0.01 & 0.01 & -1.57 \\ 
  NMF-ALS& -0.05 & 0.02 & -2.23 \\ 
  InfoMax ICA & -0.03 & 0.001 & -9.69 ***\\ 
    EBM ICA& -0.01 & 0.01 & -1.66 \\ 
  JADE ICA& -0.01 & 0.001 & -3.54 ***\\ 
 K-SVD & 0.01 & 0.01 & 0.54 \\ 
  NMF-PG & 0.07 & 0.02 & 3.25 ***\\ 
  $L1$ Regularized Learning & 0.07 & 0.001 & 13.99 ***\\ 
   \hline
\end{tabular}
\caption{Encodings using spatial maps with high intensity in CSF regions had reduced classification accuracy, while spatial maps with variable extractions in white-matter and grey-matter regions had higher classification accuracy.  Baseline is set to Fast ICA classification accuracy.  Scan-ID and Subject-ID were included as random effects within a general linear mixed-effects regression model to adjust for multiple comparisons.  *** = p $<$ .001  \label{tab:physioclass}}
\end{table}

\section{Discussion}
The BSS algorithms can be formulated as theories of functional organization and encoding. Comparing the classification accuracy can neither validate nor invalidate any specific hypothesis, but the success of a specific BSS algorithm provides evidence for a unique biological interpretation of generative activity.  Interpreting features in any decoding scheme is complex (\cite{guyon2003introduction}). As recently highlighted by \cite{haufe2014interpretation}, feature weights should not be interpreted directly. When features have a shared covariance structure, irrelevant features can be assigned a strong weight to compensate for shared noise in feature space. However, as these authors also point out, when the features are independent, the shared covariance is minimized and the interpretation theoretically becomes more tractable. 

Our experiments showed that algorithms which enforced sparsity, instead of merely encouraging it, frequently had the highest classification accuracy compared to the independence and sparse coding algorithms. When we implemented K-SVD with an overcomplete basis of 300 networks, the classification accuracy remained similar.  The non-negativity algorithms had the least accurate classification accuracy. Among the ICA algorithms, Fast ICA had the highest classification accuracy.  These trends were consistent for the extraction of 20 and 50 networks, and for different levels of data cleaning (removing artifactual components).  For all algorithms, extracted spatial maps containing more regions of CSF led to worse classification accuracy ($p<0.001$). CSF and white matter artifacts were purposefully removed during the artifact cleaning, so this CSF measures the residual noise.  Algorithms which showed large variability in their extraction of white matter regions had significantly higher classification accuracy ($p<0.001$).  This may suggest that purposefully selecting white matter regions to construct functional networks helps to improve classification accuracy.

\begin{figure}[h]
\begin{center}
\includegraphics[width=1.0\textwidth]{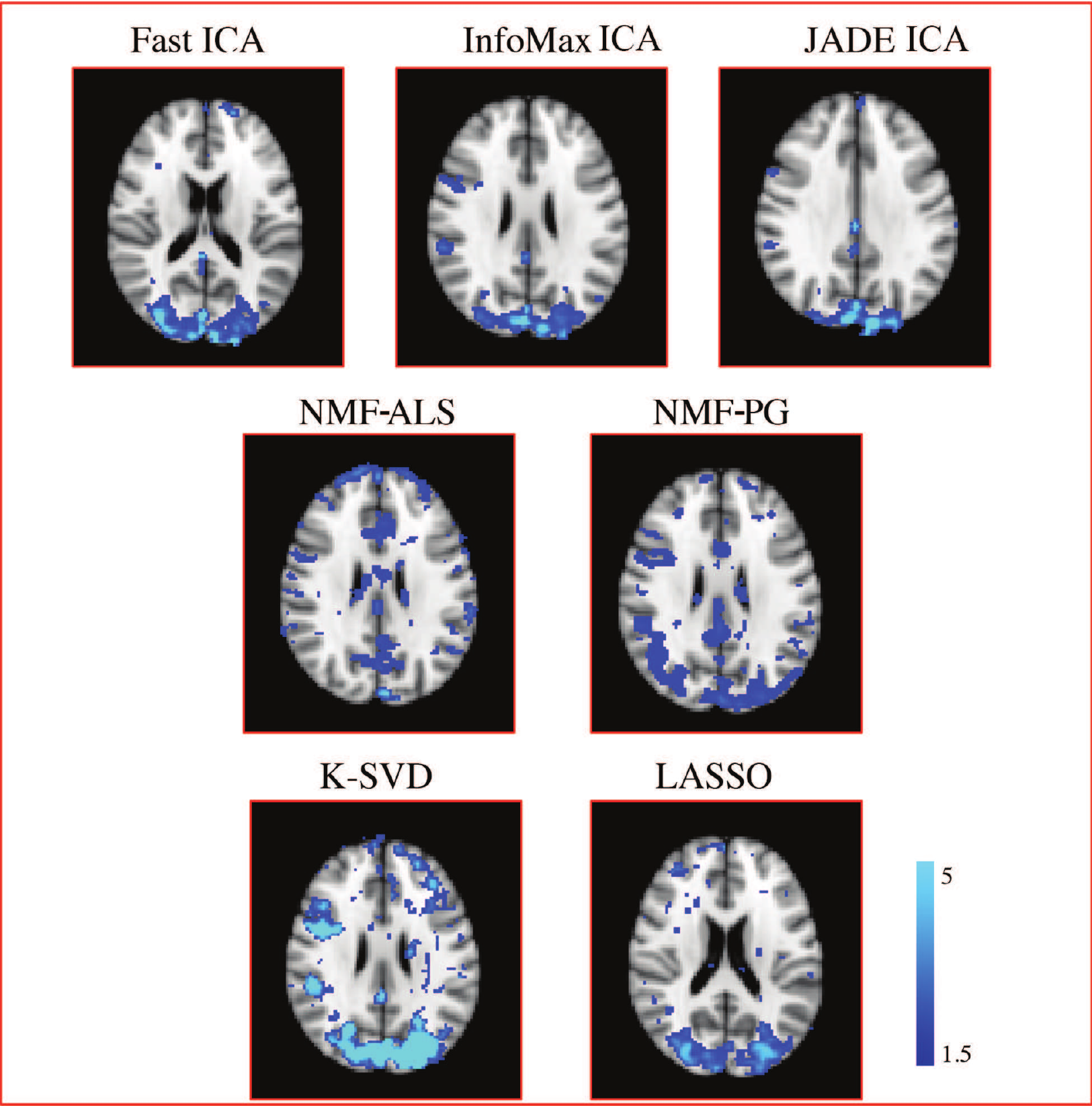}
\caption{Visual network spatial maps for each of the compared algorithms, shown for a representative subject.   Visual networks were identified via manual inspection out of 20 possible networks extracted. Values for the NMF algorithms vary from ICA and K-SVD maps, and were therefore adjusted to meet this range prior to thresholding. \label{fig:visnetfrombasismethods}}
\end{center}
\end{figure}

\begin{figure}[h]
\begin{center}
\includegraphics[width=1.0\textwidth]{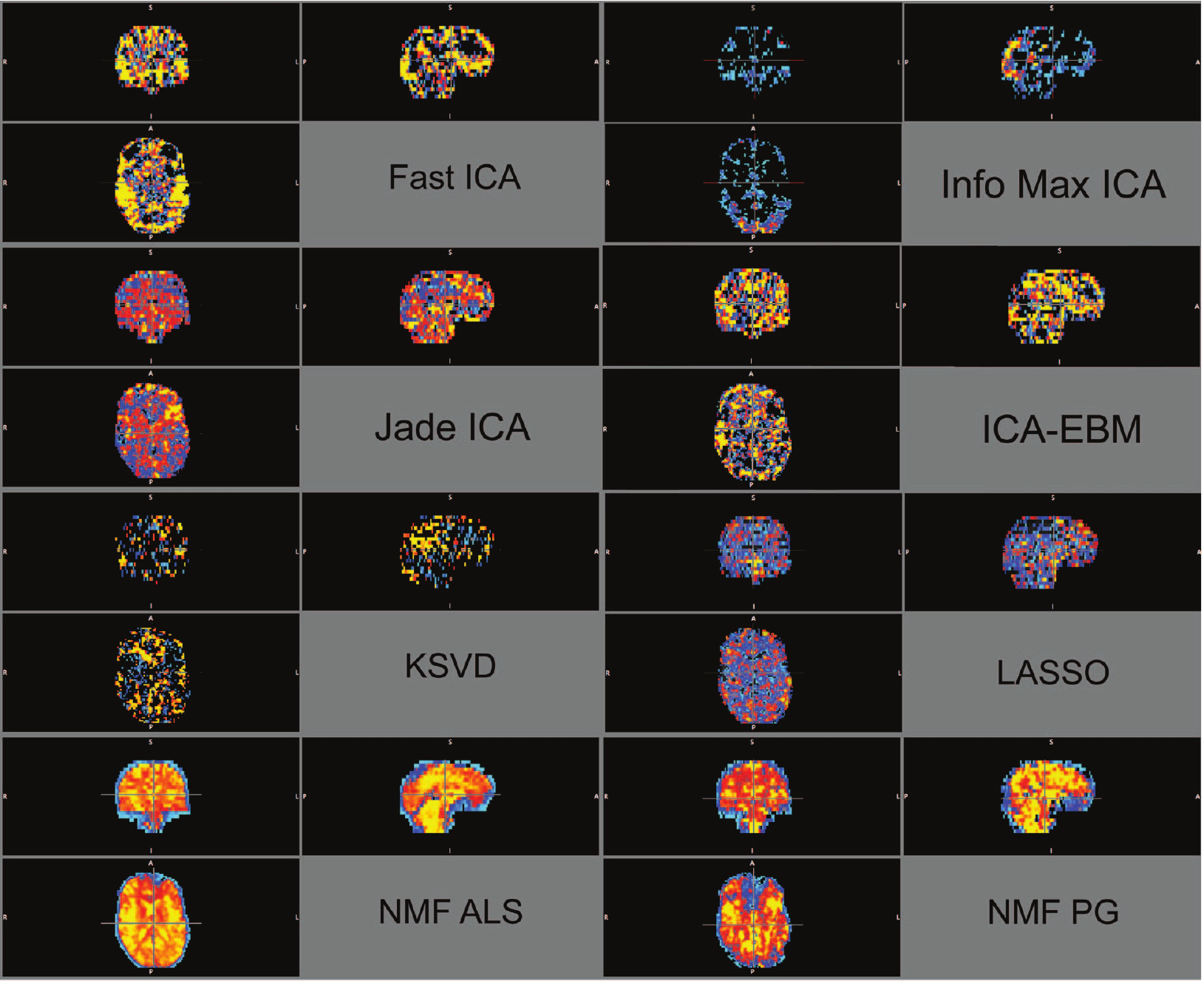}
\caption{We identified the network most associated with performing (or not) any task, using the maximal absolute correlation between the timecourse of the network and the any task/rest paradigm.  This empirically-identified network may correspond to the default mode network, since it is not task-specific (associated with only video or an audio stimuli).  The maximally correlated network for each algorithm is shown unthresholded, but consistently colored within each algorithm. High intensity regions are red, and low intensity regions are blue.  The units of the color scale depend on the algorithms; for example, NMF values would be bounded below by 0, while the intensity of the ICA spatial maps were largely centered between (-3, 3). Results shown are for the 20 dictionary elements on traditionally preprocessed data, extracted from the same scan. \label{fig:mapsfrombasismethods}}
\end{center}
\end{figure}

In Figure \ref{fig:visnetfrombasismethods}, we show thresholded spatial maps derived from the visual task for a representative subject.  These maps were all rescaled to match across algorithms.  Most of the algorithms clearly isolated the visual network.  However, the NMF algorithms resulted in weakly identifiable visual networks, and unsurprisingly this class of algorithms were outperformed by the other algorithms. It is therefore reasonable to suggest that the signed values in the BOLD signal carry either important descriptive or class specific information.  In addition to identifying possible networks through manual inspection of the spatial maps, we also identified them by the time course.  The unthresholded spatial map most correlated (absolute value) with performing any task is shown in Figure \ref{fig:mapsfrombasismethods}.  This may be the most likely candidate for the default mode network since the signs of the timecourses are arbitrarily positive or negative, and the``any task" timecourse consisted of both auditory and visual stimuli which are functionally distinct- thus precluding this as a specific task-related network. High-intensity values for that algorithm are in red, while low-intensity values are in blue, but the numeric values do not map across algorithms because of the different scaling methods these algorithms employ.  The differences among algorithms is partially attributed to the physiological profiles of the spatial maps they extracted. The 20 extracted networks for all algorithms for a single scan are provided in the Appendix as NIfTI files, along with code to reproduce these results. 

\subsection{NMF and functional MRI}
Despite the promising implications of NMF, its predictive performance was underwhelming.  The non-negativity constraint suppresses all negative BOLD activations, which has controversial hypotheses behind its generation.  The poor performance of NMF may suggest that the disregarded negative activations did contain task-related signal.  The non-negativity constraint also encourages local circuits, and geographically distributed networks may have been parsed and diluted by this.  The NMF-PG algorithm actually performed better than Fast ICA after adjusting for the CSF functional loadings and the white matter variability.  This suggests that NMF algorithms may capture higher levels of noise in the fMRI activation maps.  The NMF spatial maps had a strong resemblance to structural images with the regional intensity varying depending on the tissue type, even though the learning was done on functional data.  This was not an artifact of the initialization, as the maps were initialized to be random values. It was also not an artifact of improper convergence, as running the algorithm far beyond convergence did not change the structural-map appearance of these maps.  Rather, we speculate that the likely default mode may have been suppressed in NMF, since the DMN is anticorrelated with the task related networks which may have been stronger contenders spanning both auditory and visual domains.  It is possible that NMF performed poorly because it was not able to use activation of the DMN to identify the periods of rest. Finally, the NMF-ALS algorithm had difficulty extracting a full set of 20 networks on the artifact-cleaned fMRI data, although it was able to do so consistently on the fMRI data which had received only regular pre-processing. Collectively, this suggests that NMF may find its purpose not in capturing task-related activation maps, because of its suppression of all negative signals.

\subsection{Limitations}

There are several limitations to this paper.  We assessed how brain networks are encoded using a decoding approach- however, this is an indirect and a secondary measurement of performance, complicated by an unknown ground truth.  It is unknown how many brain networks exist in a given subject, and whether these brain networks are consistent across subjects.  Our interrogation into these networks is based upon our perturbation of the system using a stimulus, where we use task decoding to identify properties of networks which should exist based on the visual and audio stimuli present.  We compared the performance of the algorithms using their prediction power in a supervised learning model.  This suggests that, out of the algorithms considered here, the sparse coding assumptions may be more biologically plausible in their accounting for task-related activity changes, but restricts our inference only to networks which are task-associated. Moreover, this does not suggest that the frameworks imposed by these specific algorithms are the best explanation for how functional networks are organized; rather, these analyses suggest that out of the linear algorithms defined by specific constraints, the sparse coding constraint may best capture task-related activations.  For the classification, given the computational considerations when optimizing many parameters, we used only a fixed 'C' penalty term.  It is possible that further hyperparameter optimization may alter results. However, in all cases the total number of features was already relatively small (20 or 50).  Given that both algorithms internally penalize for overfitting, it is unlikely that this optimization would have a large effect.  Finally, although we used 304 scans for this study, these scans originated from 51 subjects.  Although we controlled for the effects of Subject, Scan, and Session, there may have been other unknown factors which introduced covariance.

The secondary artifact-cleaning step used Fast ICA to propose possible noisy networks, which were flagged and removed before reconstructing the scans.  The ICA-based artifact cleaning may have introduced some disadvantages to the other algorithms, since these algorithms were run on ICA-cleaned data.  We cleaned the data (motion correction, etc...) in two stages, through traditional pre-processing and secondarily through artifact-removal using FIX.  Despite this, some residual noise may still have remain, as demonstrated by the CFS intensity in the cleaned data predicting poor classification accuracy within a scan.  Although we referred to the spatial maps as brain networks here, the variability of these networks for the different algorithms gives pause as to whether these networks are really the linear combinations as the BSS algorithms assume to produce functional activity.  We did not evaluate every existing variation of ICA, NMF, and other sparse coding algorithms, but relied instead on the most popular variations.  Similarly, these analyses only compared within-scan network extractions, and may not generalize to group analyses where networks are extracted across large numbers of subjects and scans. It is possible that the sparse coding algorithms may be hypertuned to the unique activity within a single scan, and that the networks observed when applied to groups may not be flexible enough to account for the variation seen across subjects. Investigating how these networks change in group analyses, and whether sparsifying methods are still superior on a group scale, are both directions for future research.

\section{Conclusion}

The most common methods of interpreting functional brain networks may not be the most effective for describing the networks active within a single scan.  More generally, these results suggest that the functional organization of the brain may be constructed better by sparse coding algorithms than by spatial independence or parts-based representations, at the lowest levels.  We argue these results are reasonable, where a sparse operational framework would support an efficient use of biological resources.  This does not preclude a network-based view of functional activity.  The SVM algorithm, at its heart, uses weighted combinations of sparse elements to predict which activity a subject was performing.  This suggests that the sparse interpretation allows a more flexible construction of the comprehensive networks than networks which by their mathematical assumptions nurture more fully-defined functional maps.

We suggest that sparsity is a reasonable developmental constraint, insomuch as redundant (non-sparse) network components are costly in energy consumption and complexity. Sparse coding schemes reduce energy consumption by minimizing the number of action potentials necessary to represent information within neural codes (\cite{spratling2014classification}; \cite{sengupta2010action}).  Moreover, the specialization and discouragement of local multitasking associated with spatial sparsity may be more reasonable than the ``boundless energy" framework of statistical independence, where a single voxel is not constrained by its overall (total) network membership. The non-negativity NMF algorithms performed poorly, suggesting that the negative BOLD signals are indeed important measures of task-related functional activity.  In ICA, a single voxel's intensity in a given network is not influenced by the separate cognitive processes (networks) to which that voxel contributes, permitting unlimited multitasking.  The success of the sparse coding algorithms suggests that, during periods of activity, local activity is specialized.

The power of sparse algorithms may be in their disbursement of functional activity across multiple networks; the components they then nominate hold power when combined with other sparse components, suggesting they form a metabasis for cognitive activity rather than complete networks.  This may indicate that these sparse elements will prove more useful for hierarchical algorithms such as deep learning methods.  This is a direction for future research. The experiments presented here suggest the importance of sparsity, but indeed sparsity may not be the key factor in decoding fMRI activity.  When including the anatomical profiles of the spatial networks, the extraction of CSF regions (a proxy measure of overall noise) overtook sparsity in explanatory power to predict classification accuracy.  This may suggest then that sparsity is a proxy measure for reduced noise, but not necessarily the driving source behind the predictive power. The sparser maps may be more successful across all algorithms because they correspond well to the actual biological processes, because they omit regions containing noise, or even because the sparser basis sets span the partitioning space better. Although sparsity presents an optimal encoding framework for applying decoding models, the mechanism underlying its success remains unknown.

\newpage
\section{Acknowledgments}
Our sincere appreciation to M.S. Cohen and M.A.O. Vasilescu for constructive feedback and insight on this manuscript.

Ariana E. Anderson, Ph.D., holds a Career Award at the Scientific Interface from BWF and is supported by R03 MH106922. Pamela K. Douglas is supported by Klingenstein Third Generation Fellowship, Keck Foundation, and NIH/National Center for Advancing Translational Science (NCATS) UCLA CTSI Grant Number UL1TR000124. Ying Nian Wu is supported by NSF DMS 1310391, ONR MURI N00014-10-1-0933, DARPA MSEE FA8650-11-1-7149.  Arthur L. Brody is supported by the National Institute on Drug Abuse (R01 DA20872), the Department of Veterans Affairs, Office of Research and Development (CSRD Merit Review Award I01 CX000412), and the Tobacco-Related Disease Research Program (23XT-0002). 
\newpage

\newpage
\section{Figures and Legends}

\end{document}